\documentclass[aps,pre,showpacs,amsmath,amssymb,longbibliography]{revtex4-2}
\usepackage{graphicx}
\usepackage{float}

\newcommand{\be}{\begin{equation}}
\newcommand{\ee}{\end{equation}}
\newcommand{\fig}[1]{Fig.~\ref{#1}}
\newcommand{\Fig}[1]{Figure~\ref{#1}}
\newcommand{\sect}[1]{Sec.~\ref{#1}}

\newcommand{\eq}[1]{Eq.~(\ref{#1})} 
\newcommand{\Eq}[1]{Equation~(\ref{#1})}

\graphicspath{ {./figures/} }

\begin{document}
	\title{Hidden scale invariance in the Gay-Berne model}
	\date{\today}
	\author{Saeed Mehri}
	\affiliation{\textit{Glass and Time}, IMFUFA, Department of Science and Environment, Roskilde University, P.O. Box 260, DK-4000 Roskilde, Denmark}
	\author{Jeppe C. Dyre}\email{dyre@ruc.dk}
	\affiliation{\textit{Glass and Time}, IMFUFA, Department of Science and Environment, Roskilde University, P.O. Box 260, DK-4000 Roskilde, Denmark}	
	\author{Trond S. Ingebrigtsen}\email{trond@ruc.dk}
	\affiliation{\textit{Glass and Time}, IMFUFA, Department of Science and Environment, Roskilde University, P.O. Box 260, DK-4000 Roskilde, Denmark}

\begin{abstract}
This paper presents a numerical study of the Gay-Berne liquid crystal model with parameters corresponding to calamitic (rod-shaped) molecules. The focus is on the isotropic and nematic phases at temperatures above unity. There we find strong correlations between the virial and potential-energy thermal fluctuations, reflecting the hidden-scale invariance symmetry. This implies the existence of isomorphs, which are curves in the thermodynamic phase diagram of approximately invariant physics. We study numerically one isomorph in the isotropic phase and one in the nematic phase. In both cases, good invariance of the dynamics is demonstrated via data for the reduced-unit time-autocorrelation functions of the mean-square displacement, angular velocity, force, torque, and first- and second-order Legendre polynomial orientational order parameters. Deviations from isomorph invariance are observed at short times for the orientational time-autocorrelation functions, which reflects the fact that the moment of inertia is assumed to be constant and thus not isomorph invariant in reduced units. Structural isomorph invariance is demonstrated from data for the radial distribution functions of the particles and their orientations. For comparison, all quantities were also simulated along an isochore of similar temperature variation in which case invariance is not observed. We conclude that the thermodynamic phase diagram of the calamitic Gay-Berne model is essentially one-dimensional in the studied regions as predicted by isomorph theory, a fact that potentially allows for simplifications of future theories and numerical studies.
\end{abstract}
\maketitle

\section{Introduction}

Liquid crystals (LCs), rod-like polymers, and disk-formed particles  all involve molecules with a high degree of shape anisotropy \cite{jurasek2017self}. LCs occur in many contexts, ranging from the well-known display applications to biological systems \cite{woltman2007liquid,de2017rod,tian2018self}. Depending on the density and the temperature, the anisotropy of LCs may lead to different structural phases. For the thermometric LCs in focus here, changing temperature and density may cause transitions from the ordinary crystalline state to smectic, nematic, and isotropic phases \cite{woltman2007liquid}.

Pure fluids and mixtures consisting of aspherical particles have been subject to many theoretical, experimental, and simulations studies \cite{antypov2004role,cifelli2006smectic,berardi2007computer,vadnais2008study}. Theoretical studies are typically based on the Fokker-Planck equation \cite{mendez2010relaxation,doi1988theory}, generalized Langevin equations \cite{kalmykov2001rotational,hernandez1996rotational}, Onsager theory \cite{szalai1998external}, density-functional theory \cite{del1996wetting}, or generalized van der Waals descriptions \cite{del1995surface}. Different numerical techniques like Monte-Carlo and molecular dynamics (MD) have been applied for studying the phase behavior, thermodynamics, structure, and dynamics of rigid anisotropic molecules forming LCs \cite{allen1996computer}.

Depending on the type of interaction between the molecules, one can classify LC models into two main groups \cite{allen1995simulations}. The first group involves models of hard particles with a non-spherical shape \cite{allen1993hard}. In such models there are no attractive interactions, i.e., the potential is purely repulsive and short-ranged. The main motivation for this approach is the success of the hard-sphere model in explaining the properties of simple liquids \cite{hansen1986theory,dyre2016simple}. Extensive simulation studies have used this approach to investigate structure and dynamics of LC fluids for different shape anisotropies (prolate ellipsoids, spherocylinders, rods, disks, etc) \cite{eppenga1984monte,allen1987observation,frenkel1987computer}. In the other main class of LC models, both short-range repulsive and long-range attractive interactions are taken into account. Several models for fluids of aspherical particles have been introduced for LC studies, e.g., the Kihara potential \cite{kihara1963convex}, the site–site potential \cite{streen1977liquids}, the Gaussian overlap model \cite{berne1972gaussian}, and the Gay–Berne (GB) model \cite{gay1981modification}. By using site-site potentials one can realistically mimic the structure of LC molecules and compare results to experiments \cite{egberts1988molecular,komolkin1989computer,wilson1991computer,wilson1992structure,paolini1993simulation,patnaik1995modelling}, but unfortunately such models usually require huge computational resources. This is why most simulations so far have been conducted for relatively small system sizes. An exception is the GB model based on the Lennard-Jones (LJ) pair interaction, which is computationally cheap and still realistic. For this reason, the GB model has become a generic LC model. The GB model gives rise to a rich mesogenic behavior \cite{tian2018self}. Previous numerical studies of this model focused on its phase behavior
\cite{adams1987computer,luckhurst1990computer,de1991location,alejandre1994molecular,chalam1991molecular,andrew1993monte,hashim1995computer}, per-particle translational and orientational dynamics \cite{de1992dynamics}, interfacial properties \cite{del1995computer}, elastic constants \cite{stelzer1995molecular}, thermal conductivity \cite{sarman1994molecular,sarman1993self}, and viscosity  \cite{sarman1993statistical,smondyrev1995viscosities}. Analytical perturbation theories have also been applied in order to explore the phase diagram of GB fluids \cite{gupta1989computer,velasco1995liquid}.

The GB model allows one to describe different shape anisotropy, spanning from elongated ellipsoids to thin disks \cite{gay1981modification}. The GB potential depends on four dimensionless parameters, which is often indicated by the notation $GB(\kappa, \kappa^{\prime}, \mu, \nu)$. The four parameters control the shape of the molecules and the strength of the interaction between them. $GB(3,5,2,1)$ is the most studied case, mimicking rod-shape molecules, and in this case the phase diagram and orientational order parameter are known \cite{de1991liquid}. Moreover, the velocity time-autocorrelation function \cite{de1992dynamics}, viscosity \cite{smondyrev1995viscosities}, elastic constants \cite{allen1996molecular}, free energies and enthalpy \cite{del1996wetting}, isotropic-nematic transition \cite{de1991location}, liquid-vapor coexistence curve \cite{rull2017computer}, stress-tensor components \cite{sarman2015non}, and self-diffusion coefficient \cite{sarman2016self} have been studied for the $GB(3,5,2,1)$ model.

This paper presents a study of the Gay-Berne model with the above parameters corresponding to calamitic, i.e., rod-shaped elongated, molecules at high temperatures, where the model is shown to obey the symmetry of hidden scale invariance. According to this symmetry, the system is expected yo have isomorphs, which are curves in the thermodynamic phase diagram along which structure and dynamics are almost invariant when given in properly reduced units. A recent study of ours showed this for the more exotic discotic Gay-Berne model $GB(0.345,0.2,1,2)$ in the isotropic phase \cite{meh22}; the present paper demonstrates the existence of isomorphs in both the isotropic and the nematic phase of a more standard Gay-Berne model. The study given below nicely confirms a recent work by Liszka and co-workers \cite{lis22}, although that paper however focused more on density-scaling (a specific consequence of isomorph theory) than on demonstrating isomorph invariance of structure and dynamics.

\section{Gay-Berne Potential}
The GB potential between pairs of particles (``molecules''), $GB(\kappa, \kappa^{\prime}, \mu, \nu)$, is characterized by four dimensionless parameters: $\kappa\equiv \sigma_e/\sigma_s$ where $\sigma_e$ and $\sigma_s$ are lengths, $\kappa'\equiv\varepsilon_{ss}/\varepsilon_{ee}$ where $\varepsilon_{ss}$ and $\varepsilon_{ee}$ are energies, while $\mu$ and $\nu$ are exponents.

The GB pair potential $v_{\rm GB}$, which is basically a direction-dependent LJ pair potential, is defined as follows

\begin{subequations}
	\label{GB_pot}
	\begin{align}
		v_{\rm GB}(\textbf{r}_{ij},\hat{\textbf{e}}_i,\hat{\textbf{e}}_j) &= 4\varepsilon (\hat{\textbf{r}}, \hat{\textbf{e}}_i, \hat{\textbf{e}}_j) \left[\left(\sigma_s/\rho_{ij}\right)^{12} - \left(\sigma_s/\rho_{ij}\right)^{6} \right],
		\label{GB_pot_a} \\
		\rho_{ij} &= r_{ij} - \sigma(\hat{\textbf{r}}, \hat{\textbf{e}}_i, \hat{\textbf{e}}_j) + \sigma_s\,.
		\label{GB_pot_b}
	\end{align}
\end{subequations}
Here, $r_{ij}$ is the distance between molecules $i$ and $j$, $\hat{\textbf{r}}\equiv\textbf{r}_{ij}/r_{ij}$ is the unit vector along the vector from molecule $i$ to molecule $j$ denoted by $\textbf{r}_{ij}$, and $\hat{\textbf{e}}_i$ and $\hat{\textbf{e}}_j$ are unit vectors along the major axes of molecules $i$ and $j$. The GB molecule is roughly an ellipsoid of two diameters $\sigma_s$ and $\sigma_e$, and one defines

\begin{subequations}
	\label{GB_sigma}
	\begin{align}
		\sigma(\hat{\textbf{r}}, \hat{\textbf{e}}_i, \hat{\textbf{e}}_j) &= \sigma_s \bigg[1-\dfrac{\chi}{2} \bigg(\dfrac{(\hat{\textbf{e}}_i\cdot\hat{\textbf{r}}+\hat{\textbf{e}}_j\cdot\hat{\textbf{r}})^2}{1+\chi(\hat{\textbf{e}}_i\cdot\hat{\textbf{e}}_j)}+ \dfrac{(\hat{\textbf{e}}_i\cdot\hat{\textbf{r}}-\hat{\textbf{e}}_j\cdot\hat{\textbf{r}})^2}{1-\chi(\hat{\textbf{e}}_i\cdot\hat{\textbf{e}}_j)}\bigg) \bigg]^{-1/2}
		\label{GB_sigma_a}\\
		\chi&=\dfrac{\kappa^2-1}{\kappa^2+1}\,.
		\label{GB_sigma_b}
	\end{align}
\end{subequations}
Physically, $\chi$ is a shape anisotropy parameter and $\kappa$ quantifies the molecular elongation. The case $\kappa=1$ ($\chi=0$) represents spherical molecules, the case $\kappa\rightarrow\infty$ ($\chi\rightarrow1$) corresponds to very long rods, and the case $\kappa\rightarrow 0$ ($\chi\rightarrow-1$) corresponds to very thin disks. The energy term is given as follows
	
\begin{subequations}
	\label{GB_epsilon}
	\begin{align}
		\varepsilon(\hat{\textbf{r}},& \hat{\textbf{e}}_i, \hat{\textbf{e}}_j) = \varepsilon_0\, 
		\left(\varepsilon_1(\hat{\textbf{e}}_i,\hat{\textbf{e}}_j)\right)^\nu
		\left(\varepsilon_2(\hat{\textbf{r}}, \hat{\textbf{e}}_i, \hat{\textbf{e}}_j)\right)^\mu
		\label{GB_epsilon_a}\\
		\intertext{in which}
		\varepsilon_1(\hat{\textbf{e}}_i,\hat{\textbf{e}}_j)&=\big(1-\chi^2(\hat{\textbf{e}}_i\cdot\hat{\textbf{e}}_j)^2\big)^{-1/2}
		\label{GB_epsilon_b}\\
		\varepsilon_2(\hat{\textbf{r}}, \hat{\textbf{e}}_i, \hat{\textbf{e}}_j)&= 1-\frac{\chi'}{2}\biggl(\dfrac{(\hat{\textbf{e}}_i\cdot\hat{\textbf{r}}+\hat{\textbf{e}}_j\cdot\hat{\textbf{r}})^2}{1+\chi'(\hat{\textbf{e}}_i\cdot\hat{\textbf{e}}_j)}+ \dfrac{(\hat{\textbf{e}}_i\cdot\hat{\textbf{r}}-\hat{\textbf{e}}_j\cdot\hat{\textbf{r}})^2}{1-\chi'(\hat{\textbf{e}}_i\cdot\hat{\textbf{e}}_j)}\biggr)
		\label{GB_epsilon_c}\\
		\intertext{and the energy anisotropy parameter is given by}
		\chi'&=\frac{\kappa'^{1/\mu}-1}{\kappa'^{1/\mu}+1}\,.
	\end{align}
\end{subequations}
The energies $\varepsilon_{ss}$ and $\varepsilon_{ee}$ are the well depths of the potential in the side-side and end-end configurations, respectively. Henceforth, unless isomorph-theory reduced units are used (see \sect{sec:isom}), $\sigma_s$ defines the length unit used and $\varepsilon_0$ the energy unit. The density $\rho$ and the temperature $T$ are always given in these units. 
	
The $GB(3,5,2,1)$ model was introduced in 1981 by Gay and Berne, inspired by the Gaussian overlap model of Berne and Pechukas \cite{berne1972gaussian,gay1981modification}. For realistic LCs the length-to-width ratio is at least $3$, leading to the choice of $\kappa=3$ by Gay and Berne \cite{gay1981modification}. To obtain the other parameters, the GB pair potential was compared to the case of a pair of linear molecules consisting of four LJ particles placed on a line such that the length-to-width ratio equals $3$. This results in $\kappa^{\prime}=5,~\mu=2$, and $\nu=1$ \cite{gay1981modification}.

As mentioned, the $GB(3,5,2,1)$ model shows a rich phase behavior with isotropic, nematic, and smectic B phases \cite{adams1987gr,luckhurst1990computer,de1991liquid}. Actually, the model has also the following phases: smectic A \cite{luckhurst1990computer}, tilted smectic B \cite{de1991liquid}, and rippled smectic B phases \cite{hashim1995computer}. In some cases, more involved versions of the GB potential have been investigated by introducing, e.g., dipolar forces \cite{satoh1996monte}, flexibility \cite{la1996rigid}, more complex shapes \cite{neal1997molecular}, or biaxial molecules \cite{cleaver1996extension}. Other sets of parameters have also been studied, and other properties have been examined, e.g., the effect of the $\nu$ exponent on the orientational order parameter \cite{mori2003brownian}, elastic constant for $GB(3,5,1,3)$ \cite{germano2002simultaneous}, diffusion coefficient in the smectic A phase of $GB(4.4,20,1,1)$ \cite{bates2004studies}, stability of the smectic phase, radial distribution function, orientational order parameter \cite{miguel2004stability}, and rotational viscosity coefficient \cite{satoh2006characteristic}. Satoh $et~al.$ studied the effect of an external magnetic field on GB fluids \cite{satoh2006molecular}. The isotropic-nematic region has been explored for different values of $\kappa$ \cite{mcdonald2000surface,huang2014calculation}. Varying $\kappa^{\prime}$ while keeping the other parameters fixed at $\kappa=3,~\mu=2$ and $\nu=1$ has been investigated in detail, the liquid-vapor region has been analyzed \cite{de1991effect,de1996effect}, and so has the equation of state, structure, and diffusion coefficient \cite{he2009self}. For discotic GB fluids the phase diagram has been obtained for different $\kappa$ and $\kappa^{\prime}$ parameters \cite{akino2001molecular,caprion2003influence,yamamoto2005brownian,meh22}. The most studied discotic model is $GB(0.345,0.2,1,2)$ \cite{cienega2014phase}, which incidentally led to an improvement of the angle of view of liquid crystals displays \cite{bushby2011liquid}.	

\begin{figure}[!h]
  \includegraphics[width=0.3\textwidth]{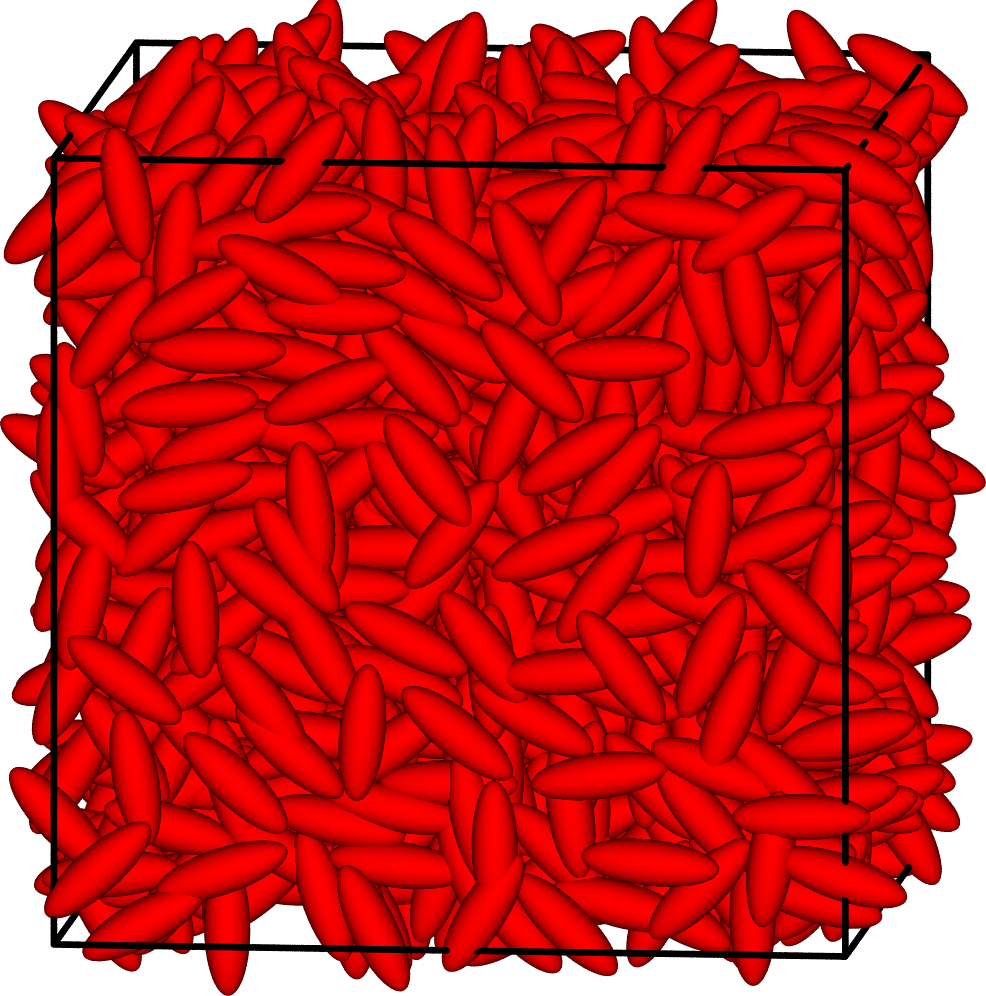}
  \includegraphics[width=0.3\textwidth]{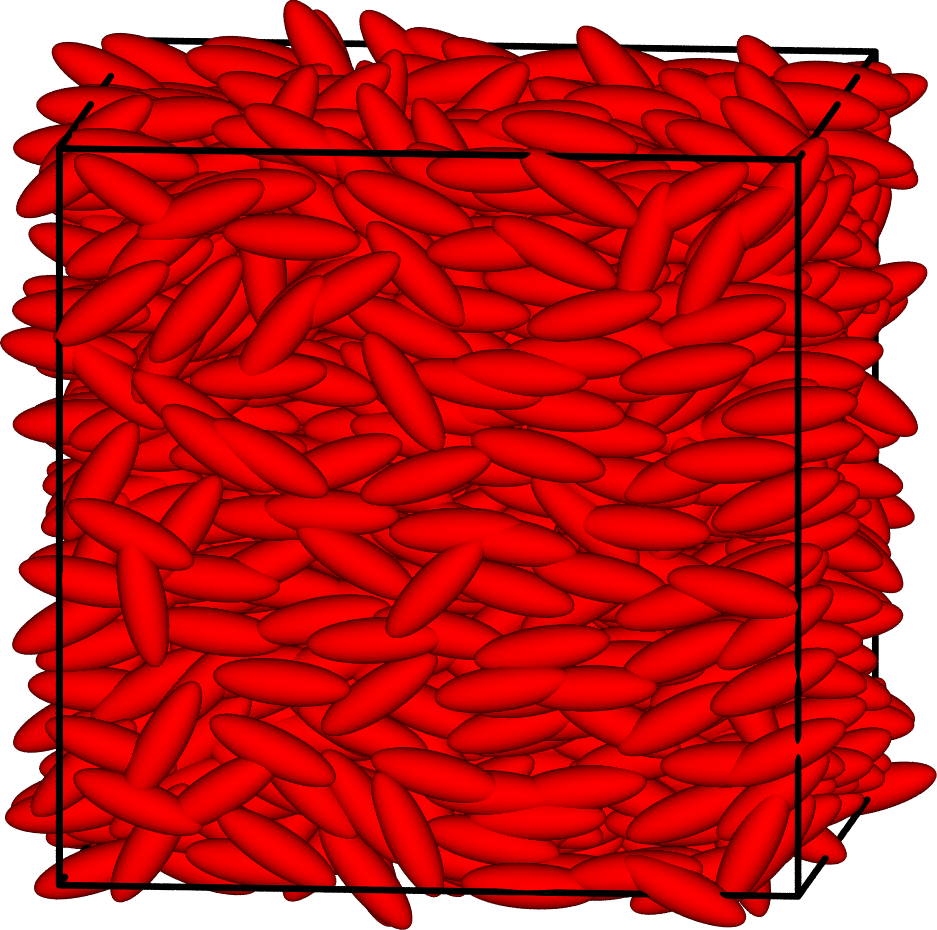}
  \includegraphics[width=0.27\textwidth]{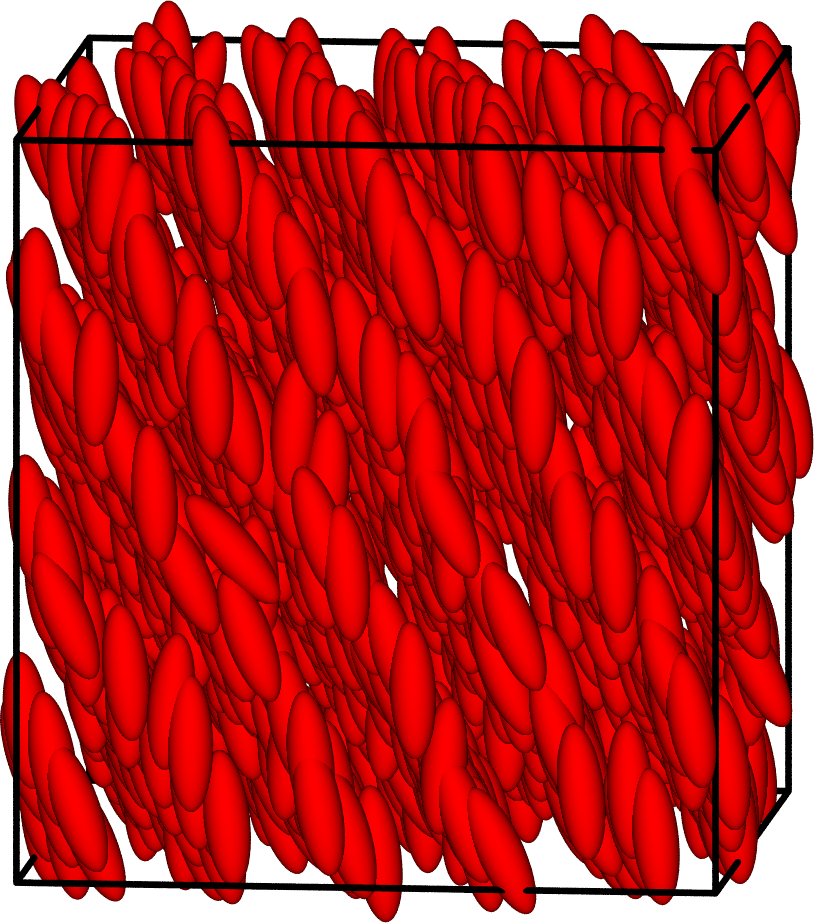}
  \caption{Snapshots of the calamitic GB model $GB(3,5,2,1)$ at three state points. 
  (a) shows the isotropic liquid phase at the state point $(\rho, T ) = (0.27, 1.2)$; 
  (b) shows the nematic phase at $(\rho, T ) = (3.3, 1.2)$; 
  (c) shows the smectic phase at $(\rho, T ) = (3.9, 1.2)$.  }  \label{fig:rod_conf}
\end{figure}

In this work we study the $GB(3,5,2,1)$ model because, as already mentioned, its phase diagram, structure, and dynamics are known. Fig. \ref{fig:rod_conf} presents snapshots of the system at equilibrium in the isotropic, nematic, and smectic phases. There is no positional or orientational ordering in the isotropic phase. In the nematic phase there is no positional ordering, but some long-range orientational ordering. In the smectic phase the molecules form parallel layers with a robust orientational ordering within the layers.

\section{R-simple systems and isomorphs}\label{sec:isom}

Recalling that the virial $W$ quantifies the part of the pressure $p$ deriving from molecular interactions via the defining identity $pV=Nk_BT+W$ (in which $V$ is the volume and $N$ is the number of molecules) \cite{bailey2013statistical}, liquids and solids may be classified according to the correlation between the equilibrium fluctuations of the virial and the potential energy $U$ \cite{ingebrigtsen2012simple}. The so-called R-simple systems, which are those with strong such correlations, are particularly simple because the thermodynamic phase diagram is basically one-dimensional instead of two-dimensional in regard to structure and dynamics \cite{ingebrigtsen2012isomorphs,ingebrigtsen2012simple,dyre2014hidden,dyre2018perspective}.

Isomorph theory dealing with R-simple systems was developed over the last decade \cite{bailey2008pressure, gnan2009pressure, veldhorst2014scaling, costigliola2016freezing}. The $WU$ Pearson correlation coefficient is defined by

\begin{equation}
	R(\rho,T)=\frac{\langle \Delta W \Delta U \rangle}{\sqrt{\langle (\Delta W)^2 \rangle \langle (\Delta U)^2 \rangle}}\,.
	\label{pearson}
\end{equation}
Here the angular brackets denote $NVT$ ensemble averages, $\Delta$ is the deviation from the equilibrium mean value, and $\rho$ is the density. Many systems, including the LJ fluid, have strong $WU$ correlations in the liquid and solid phases, whereas $R(\rho,T)$ usually decreases significantly below the critical density \cite{bel19a}. A system is considered to be R-simple whenever $R>0.9$ at the state points in question.
\begin{figure}[!h]
	\includegraphics[width=.5\textwidth]{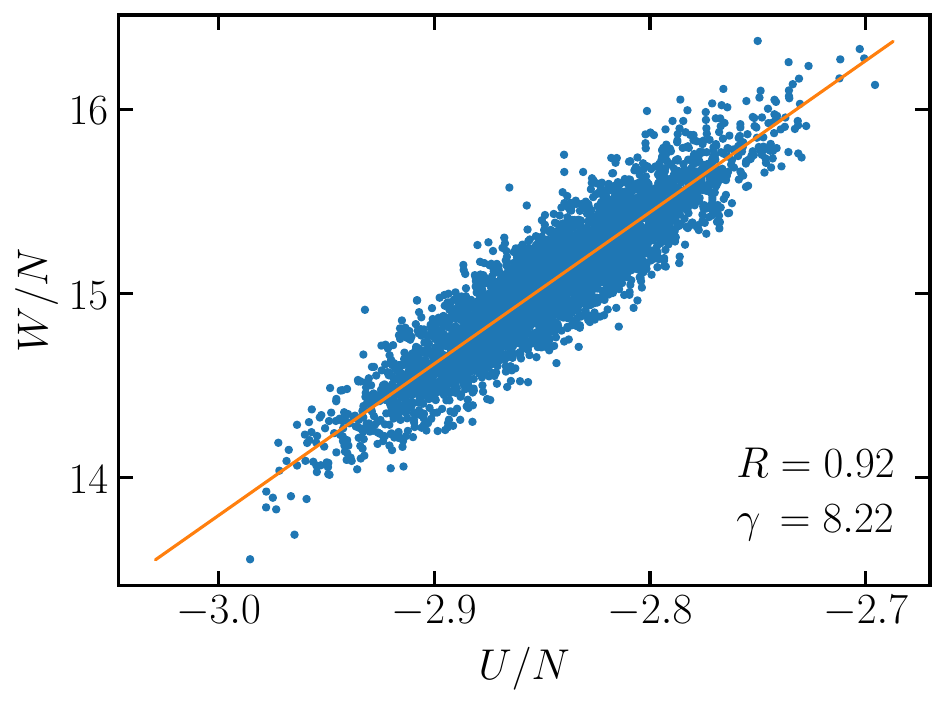}
	\caption{Scatter plot of $WU$ correlations for the $GB(3,5,2,1)$ model at the state point $(\rho,T)=(0.33, 1.2)$. The system is here strongly correlating with $R=0.92$; the density-scaling exponent $\gamma$ is $8.22$.}\label{U_W}
\end{figure}
The density-scaling exponent $\gamma$, which is characterized by $\Delta U\cong\gamma \Delta W$, is found from linear-regression fits to a $WU$ scatter plot as shown in Fig.\ref{U_W} or using the defining equation

\begin{equation}
	\gamma=\frac{\langle \Delta W \Delta U \rangle}{\langle (\Delta U)^2 \rangle}\,.
	\label{gamma}
\end{equation}
R-simple systems have curves in the phase diagram along which structure and dynamics are approximately invariant, and these curves are termed \textit{isomorphs}. It is important to emphasize that isomorph invariance only applies when data are presented in so-called reduced units. In the system of reduced units, which in contrast to ordinary units is state-point dependent, the density $\rho \equiv N/V$ defines the length unit $l_0$, the temperature defines the energy unit $e_0$, and the density and thermal velocity define the time unit $t_0$:

\begin{equation*}
	l_0=\rho^{-1/3},~~ e_0=k_{\rm B}T,~~ t_0=\rho^{-1/3}\sqrt{m/k_{\rm B}T}\,.
\end{equation*}
Here $m$ is the molecule mass. Quantities given in isomorph-theory reduced units are marked with a tilde.

Strong virial potential-energy correlations arise whenever the hidden-scale-invariance symmetry applies. This is the condition that the potential-energy ordering of same-density configurations is maintained under a uniform scaling of all coordinates \cite{schroder2014simplicity}, which is formally expressed as follows:

\begin{equation}\label{HSI}
	U(\mathbf{R}_a)<U(\mathbf{R}_b)\Rightarrow U(\lambda \mathbf{R}_a)<U(\lambda \mathbf{R}_b)
\end{equation}
where $\lambda$ is scaling factor. Consider two configurations with same potential energy, i.e., $U(\mathbf{R}_a)=U(\mathbf{R}_b)$. After a uniform scaling one has by \eq{HSI} $U(\lambda \mathbf{R}_a)=U(\lambda \mathbf{R}_b)$. By taking the derivative of this with respect to $\lambda$ one easily derives $W(\mathbf{R}_a)=W(\mathbf{R}_b)$ \cite{schroder2014simplicity}; thus same potential energy implies same virial, i.e., 100\% correlation between $W$ and $U$. \Eq{HSI} only applies approximately for realistic systems, however, so in practice one observes strong but not perfect virial potential-energy correlations.

It can be shown that \eq{HSI} implies that the reduced-unit structure and dynamics are invariant along the lines of constant excess entropy, which are by definition the isomorphs \cite{schroder2014simplicity}. Recall that a system's entropy $S$ can be expressed as that of an ideal gas plus a term deriving from the intermolecular interactions, $S=S_{\rm id}+S_{\rm ex}$. For an ideal gas one has $S_{\rm ex}=0$, for all other systems $S_{ex}<0$ because these are less disordered than an ideal gas. 

Along an isomorph one has 

\begin{equation}
	dS_{\rm ex}= \left(\frac{\partial S_{\rm ex}}{\partial T}\right)_V dT + \left(\frac{\partial S_{\rm ex}}{\partial V}\right)_T dV = 0\,.
	\label{ds}
\end{equation}
Using Maxwell's volume-temperature relation for the configurational degrees of freedom, $\left( {\partial S_{\rm ex}}/{\partial V} \right)_T=\left(  {\partial\left( W/V \right)}/{\partial T}\right)_V$, we can rewrite the \eq{ds} as

\begin{equation}
	\left( \frac{\partial S_{\rm ex}}{\partial T} \right)_V T\,d\ln T = \left( \frac{\partial W}{\partial T} \right)_V d\ln \rho\,.
\end{equation}
Using $dU=TdS_{\rm ex}-(W/V)dV$ leads to

\begin{equation}
	\left( \frac{\partial U}{\partial T} \right)_V d\ln T = \left( \frac{\partial W}{\partial T} \right)_V d\ln \rho\,,
\end{equation}
which via the fluctuation relations $\left( {\partial W}/{\partial T} \right)_V=-\langle \Delta W \Delta U \rangle/k_BT^2$ and
$\left( {\partial U}/{\partial T} \right)_V=-\langle (\Delta U)^2 \rangle/k_BT^2$ for $\gamma$ leads to the above \eq{gamma}, 

\begin{equation}
	\gamma\equiv\left( \frac{\partial \ln T}{\partial \ln \rho} \right)_{S_{\rm ex}}=\frac{\langle \Delta W \Delta U \rangle}{\langle (\Delta U)^2 \rangle}\,.
	\label{iso_gamma} 	
\end{equation}
\Eq{iso_gamma} is completely general  \cite{gnan2009pressure}. This equation is of particular interest, however, when the system has isomorphs because the equation can then be used for tracing out isomorphs without knowing the equation of state, which is done as follows. At a given state point $(\rho_1,T_1)$ one first calculates $\gamma$ from the equilibrium fluctuations of the potential energy and virial. Then, by scaling the system to a slightly different density $\rho_2$ and numerically calculating $\left( {\partial \ln T}/{\partial \ln \rho} \right)_{S_{\rm ex}}$ from \eq{iso_gamma}, one predicts the temperature $T_2$ with the property that $(\rho_2,T_2)$ is on the same isomorph as $(\rho_1,T_1)$. In the simulations of this paper we used fourth-order Runge-Kutta integration to generate isomorphs \cite{att21} involving density step sizes of approximately 1\%.

\section{Properties studied}

We simulated a system of $1372$ particles. The pair potential was cut and shifted at $r_c=4.0$ and the time step was $\Delta t = 0.001$. Because of the shape anisotropy, supplementing the standard $NVT$ Nose-Hoover algorithm for the center-of-mass motion we used the IMP algorithm for the rotational motion \cite{fincham1984more}. Different thermostats were applied for translational and rotational motion (we eventually concluded that using a single thermostat did not result in any noticeable differences, however). The molecular moment of inertia was set to $I=1$. At each simulated state point, 20 million time steps were taken to equilibrate the system before the production runs, each of which involved 67 million time steps. As a consistency check of our GB implementation, we compared the simulation results with those of the literature and found good agreement in all cases. The quantities evaluated in these comparisons, which are all defined below, were the following (where we also list the reference(s) to which data were compared): the radial distribution function $g(r)$ \cite{bates1999computer,de1991liquid}, the radial distribution orientational correlation function
$G_2(r)$ \cite{de1991liquid,de1991location,adams1987computer}, the $S_2$ orientational order parameter \cite{de1991liquid,de1991location,mendez2019equation}, and various time-autocorrelation functions \cite{de1992dynamics,jose2006multiple}.

An order parameter is a physical quantity that distinguishes between two phases. We proceed to define the second-rank orientational order parameter $S_2$ that quantifies how much the molecular orientations vary throughout the system \cite{allen2017computer}. For a uniaxial phase, $S_2$ is defined as the following sum over all molecules

\begin{equation}
    S_2 = \left\langle \frac{1}{N} \sum_i P_2(\hat{\mathbf{e}}_i\cdot \hat{\mathbf{e}}_d) \right\rangle\,.
    \label{eq:S2_1}
\end{equation}
Here $P_2$ is the second-order Legendre polynomial, $\hat{\mathbf{e}}_d$ is the director of the phase, and the angular brackets denote a time or ensemble average. This quantity takes values between 0 and 1; for a perfectly aligned system $S_2=1$ whereas $S_2=0$ implies an isotropic system.

In a simulation $\hat{\mathbf{e}}_d$ is unknown. Here the order parameter can be evaluated by maximizing $S_2$ with respect to $\hat{\mathbf{e}}_d$, which is done by rewriting \eq{eq:S2_1} as follows \cite{eppenga1984monte,allen2017computer}

\begin{equation}
    S_2=\langle \hat{\mathbf{e}}_d \cdot \mathbf{Q} \cdot \hat{\mathbf{e}}_d \rangle\,
    \label{eq:S2_2}\,.
\end{equation}
If $\otimes$ denotes a tensor product and $\mathbf{I}$ is the unity matrix, $\mathbf{Q}$ is defined by

\begin{equation}
    \mathbf{Q}= \frac{1}{2N}\sum_i (3\hat{\mathbf{e}}_i \otimes \hat{\mathbf{e}}_i -\mathbf{I})\,.
\end{equation}
It can be shows that $S_2$ is the largest eigenvalue, $\lambda_{\rm max}$, of the $\mathbf{Q}$ tensor.

\begin{figure}[!h]
  \includegraphics[width=0.5\textwidth]{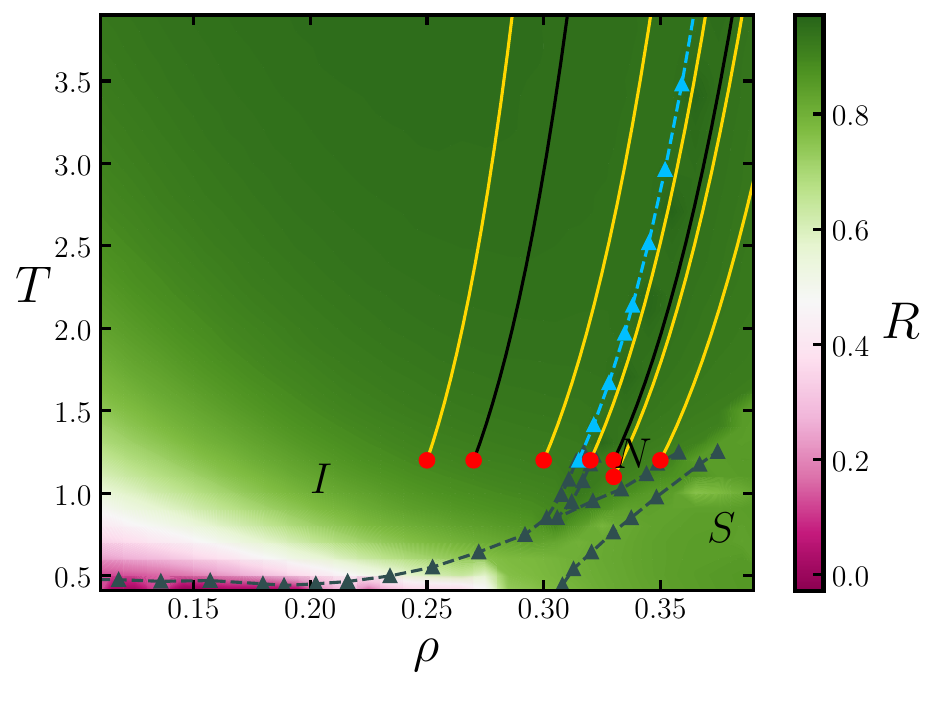}
  \includegraphics[width=0.5\textwidth]{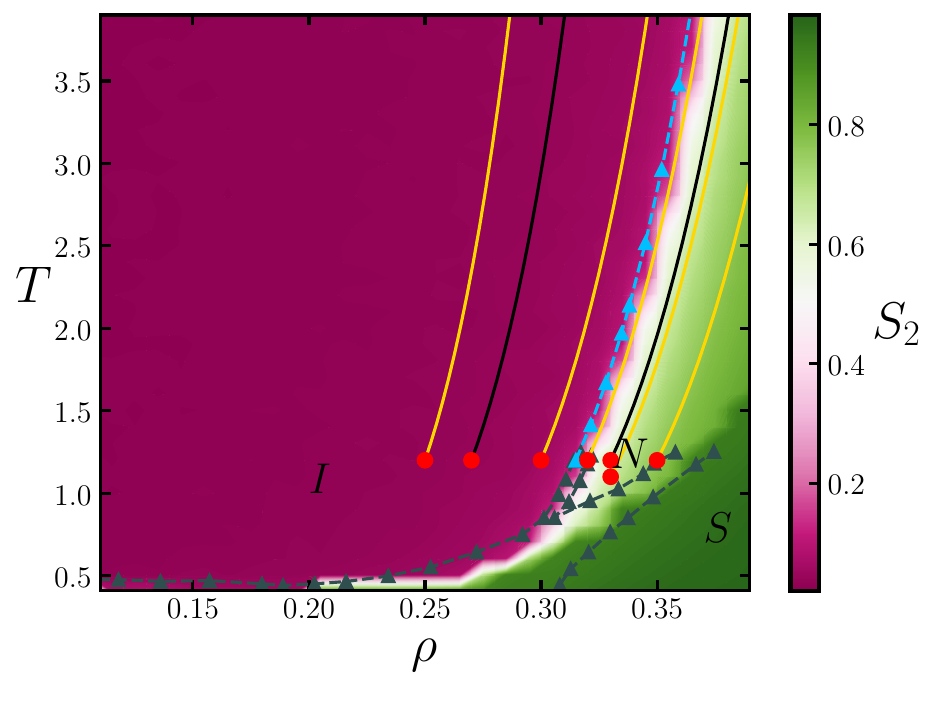}
  \caption{Density-temperature phase diagram of the $GB(3,5,2,1)$ model with (a) showing the virial potential-energy correlation coefficient $R$ (\eq{pearson}) and (b) showing the second-order orientational order parameter $S_2$ (\eq{eq:S2_2}). Dark green triangles connected by dark green dashed lines delimit the phase boundaries \cite{de2002global}. $I$ stands for the isotropic, $N$ for the nematic, and $S$ for the smectic phase. The solid yellow curves are isomorphs not investigated further here (results for these are presented in Ref. \onlinecite{Saeed_thesis}); the two black curves are isomorphs for which results are reported below, one in the isotropic phase and one in the nematic phase. The light blue triangles mark the isomorph that continues the I-N phase boundary numerical data of Ref. \onlinecite{de2002global}. The  isomorphs were determined by numerical integration of \eq{ds} starting from the following reference state points marked by the red filled circles: $(\rho_{\rm ref},T_{\rm ref})=(0.25,1.2)$, $(\rho_{\rm ref},T_{\rm ref})=(0.27,1.2)$, $(\rho_{\rm ref},T_{\rm ref})=(0.30,1.2)$, $(\rho_{\rm ref},T_{\rm ref})=(0.32,1.2)$, $(\rho_{\rm ref},T_{\rm ref})=(0.33,1.1)$, $(\rho_{\rm ref},T_{\rm ref})=(0.33,1.2)$, and $(\rho_{\rm ref},T_{\rm ref})=(0.35,1.2)$.}
  \label{fig:heatmap}
\end{figure}

\Fig{fig:heatmap} shows a ``heat-map'' phase diagram of the $GB(3,5,2,1)$ model with respect to the virial potential-energy correlation coefficient $R$ and the orientational order parameter $S_2$. By definition, the regions with $R>0.9$ are R-simple; this is where one expects isomorph theory to apply. This is not a sharp distinction, however, and many systems with $R$ between 0.8 and 0.9 have also been found to have good isomorphs. In Fig. \ref{fig:heatmap} $I$ stands for the isotropic, $N$ for the nematic, and $S$ for the smectic phase; regions in-between are those of coexisting phases. The dark green triangles marking the phase boundaries were extracted from Ref. \onlinecite{de2002global}. Selected isomorphs are marked as solid yellow lines, each of which starts from a ``reference state point'' marked as a red full circle. The main paper presents results for two isomorphs (black): one in the isotropic phase and one in the nematic phase. Results for the remaining five isomorphs are reported in Ref. \onlinecite{Saeed_thesis}. 

We conclude from \fig{fig:heatmap}(a) that there are strong correlations whenever the temperature is above unity (roughly). The isomorph reference state points were selected to obey $R>0.9$; from here on $R$ increases when density and temperature are increased along each isomorph. The Appendix give details of the two isomorphs studied by listing for each isomorph several state points and the corresponding values of $R$ and $\gamma$. Note that invariance of the physics along the isomorphs -- the main focus of this paper -- is manifested already in \fig{fig:heatmap}(b) as regards the orientational ordering because $S_2$ is clearly approximately isomorph invariant. In particular, the phase boundary approximately follows an isomorph \cite{IV,ped16}. Based on this, the light blue triangles mark the expected isotropic-nematic phase boundary. This gives an example of how isomorph theory may be used for estimating the phase boundary by allowing one to go beyond the numerical phase-boundary data of Ref. \onlinecite{de2002global} without having to perform extensive additional simulations.

Having established the phase diagram of the $GB(3,5,2,1)$ model, we next define the quantities studied. We probed the system's dynamics at the different state points by calculating the mean-square displacement (MSD) as a function of time, as well as time-autocorrelation functions defined by

\begin{equation}
	\phi_A(t)=\langle \mathbf{A}(t_0) \cdot \mathbf{A}(t_0+t) \rangle.
	\label{autocorr}
\end{equation}
Here $\mathbf{A}(t)$ is a vector or scalar molecular property and the angular brackets denote an ensemble and particle average. We evaluate below \eq{autocorr} from simulations for $\mathbf{A}$ equal to velocity, angular velocity, force, and torque. We also study the first- and second-order orientational order-parameter time-correlation function defined by

\begin{equation}
	\phi_l(t)= \langle P_l(\hat{\mathbf{e}}_i(t_0) \cdot \hat{\mathbf{e}}_i(t_0+t)) \rangle\,,
	\label{reor}
\end{equation}
in which $P_l$ is a Legendre polynomial ($l=1~\text{and}~2$). To quantify the structure we measured the standard radial distribution function, $g(r)$, as well as the radial-distribution orientational correlation function defined by

\begin{equation}
	G_2(r) \equiv \langle P_2(\hat{\mathbf{e}}_i \cdot \hat{\mathbf{e}}_j) \rangle\,
	\label{G2}
\end{equation}
where the brackets imply an average over all pairs of molecules $i$ and $j$ that are the distance $r$ apart.

\section{Results}

This section investigates to which degree the reduced-unit structure and dynamics are invariant along two isomorphs. We present data for one isomorph in the isotropic phase and one in the nematic phase (the black lines in \fig{fig:heatmap}). In realistic models isomorph invariance is only approximate, so in order to put the findings into perspective we compare the results for each isomorph with results for the isochore defined by the reference-state-point density (red points in \fig{fig:heatmap}) with the same temperature variation as that of the isomorph. For the isotropic-phase isomorph the reference state point is $(\rho_{\rm ref},T)=(0.27,1.2)$; here we cover a density variation of about $40\%$ with temperatures in the range $1.2<T<27$. For the nematic-phase isomorph the reference state point is $(\rho_{\rm ref},T)=(0.33,1.2)$; here the density varies by about $35\%$ with temperatures in the range $1.2<T<16$.

\begin{figure}[h]
    \includegraphics[width=0.45\textwidth]{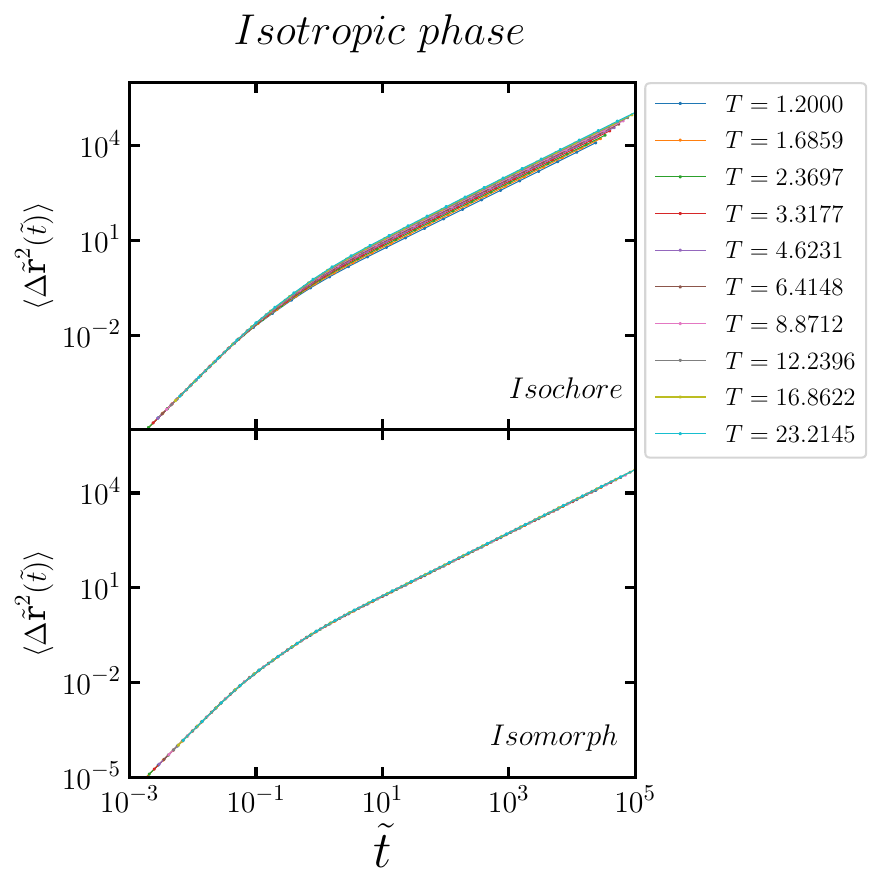}
    \includegraphics[width=0.45\textwidth]{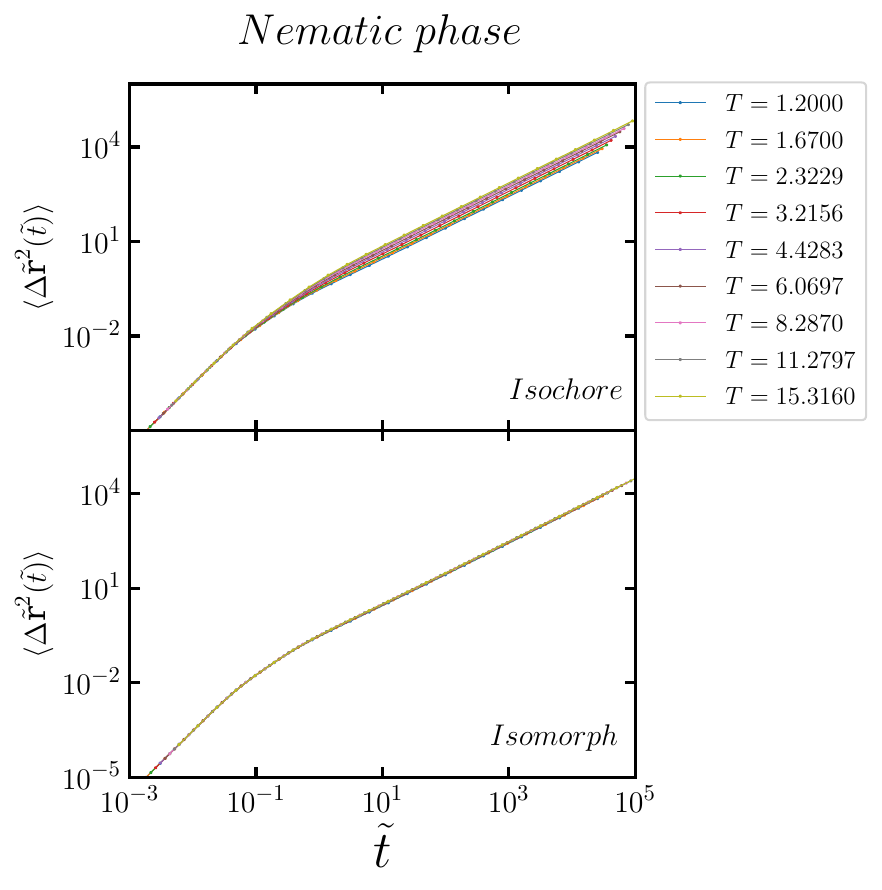}
    \caption{\label{fig:r027_T12msd} Reduced mean-square displacement as a function of the reduced time $\tilde{t}$ along an isochore and  isomorph in the isotropic phase (left) and in the nematic phase (right). In both cases, the data collapse to a good approximation along the isomorph but not along the isochore. }
\end{figure}

Fig. \ref{fig:r027_T12msd} provides data for the reduced-unit MSD along the isochore and the isomorph in the isotropic and nematic phases, respectively. The level of invariance in the center-of-mass dynamics is clearly higher along the isomorph than along the isochore. At long times, the MSD is proportional to time and the diffusion coefficients may be extracted from these data. 
\begin{figure}
    \centering
    \includegraphics[width=0.45\textwidth]{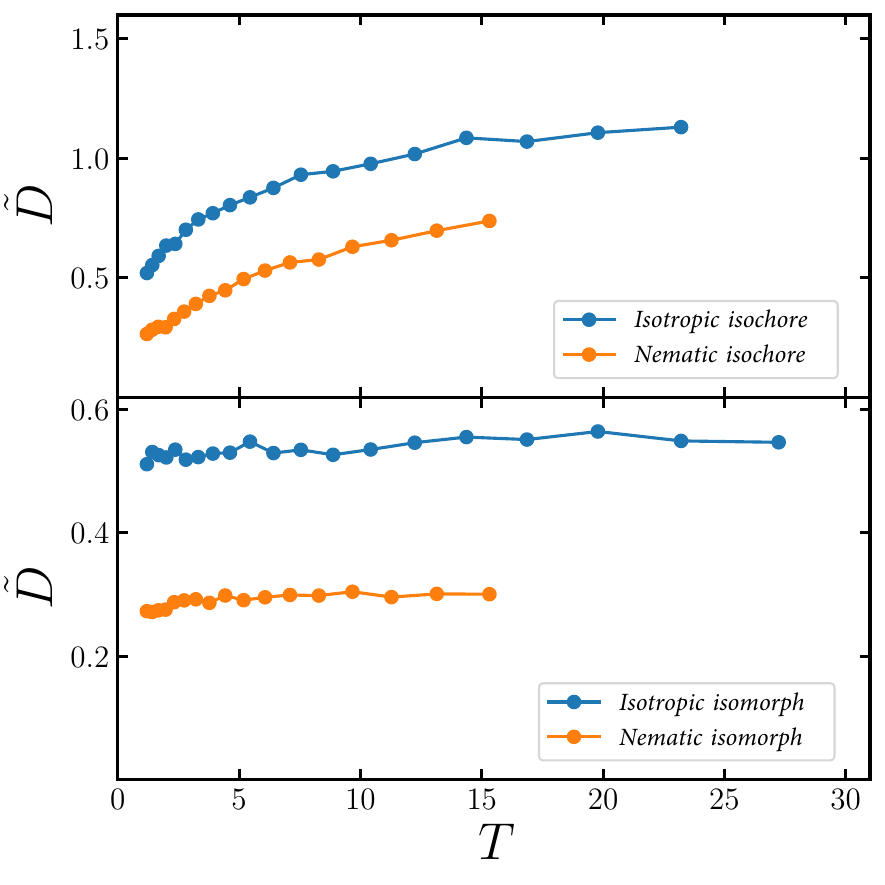}
    \caption{Reduced diffusion coefficient as a function of temperature along the isochores (upper panel) and the isomorphs (lower panel). Approximate invariance in the latter case is clearly visible.}
    \label{fig:D}
\end{figure}
\Fig{fig:D} shows the reduced diffusion coefficient as a function of temperature along the isochores and isomorphs -- approximate isomorph invariance is again clearly visible.

\begin{figure}[!h]
    \includegraphics[width=0.45\textwidth]{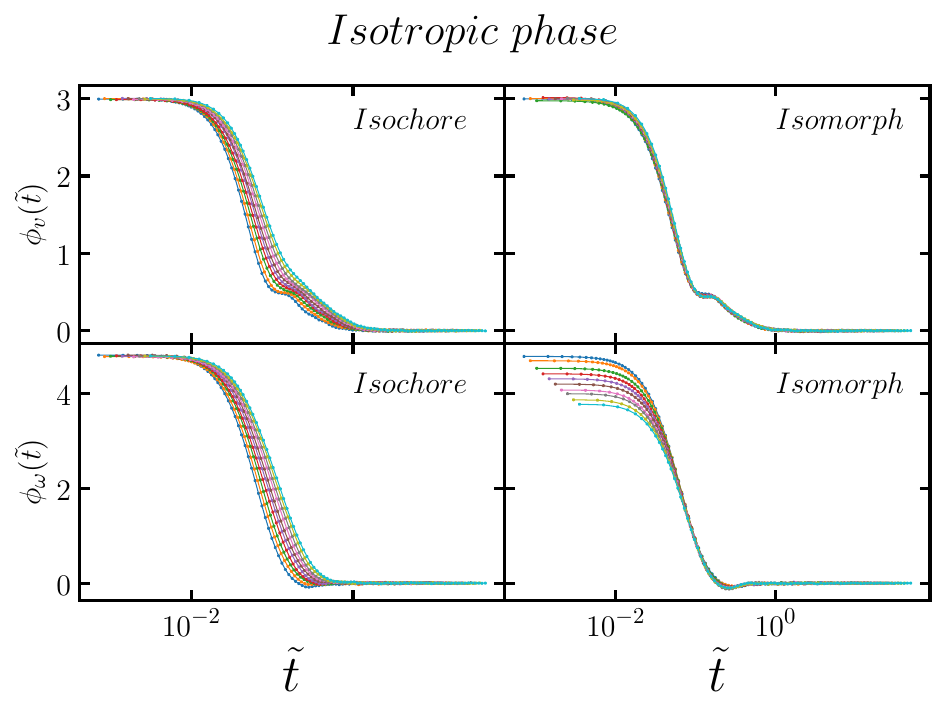}
    \includegraphics[width=0.45\textwidth]{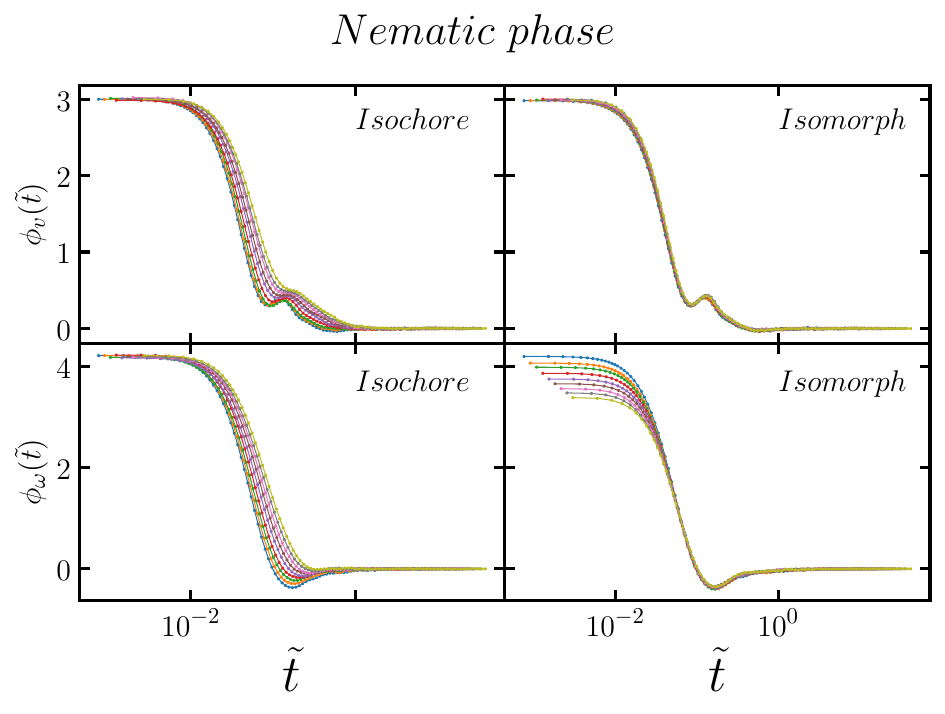}
    \includegraphics[width=0.45\textwidth]{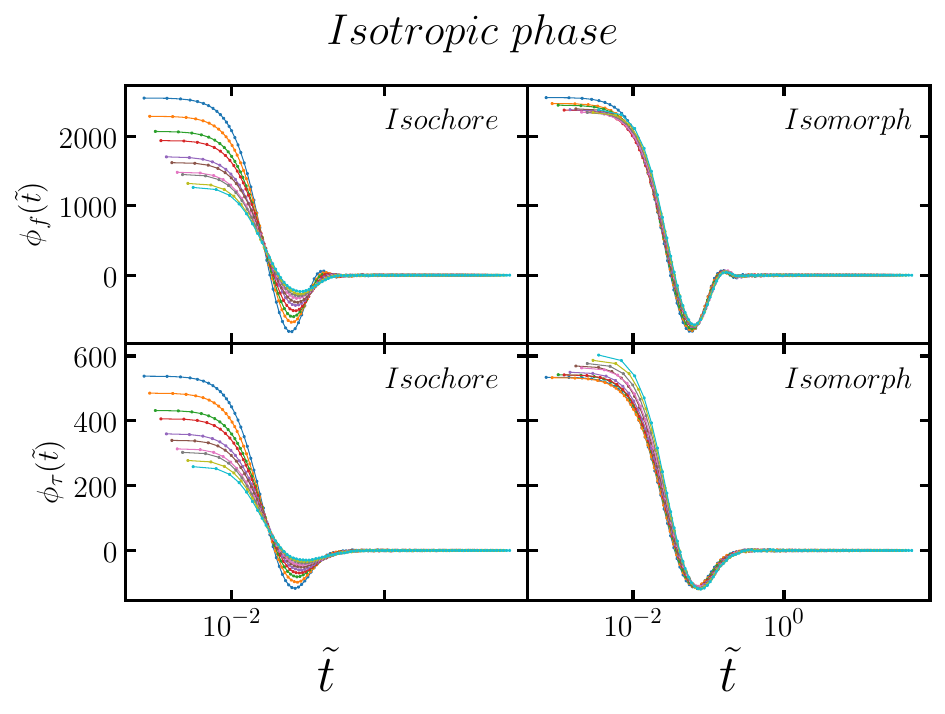}
    \includegraphics[width=0.45\textwidth]{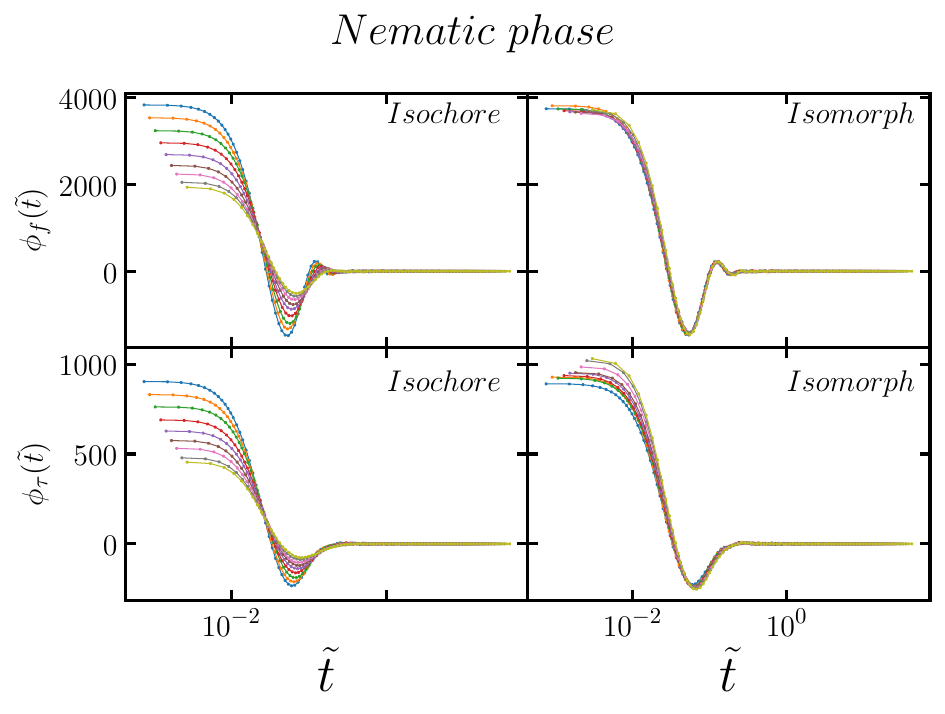}
    \caption{The upper figures show the reduced-unit time-autocorrelation functions of the velocity $v$ and the angular velocity $\omega$ along the isochore and the isomorph in the isotropic (left) and the nematic phases (right). The lower figures show the analogous results for the reduced-unit time-autocorrelation functions of the force $f$ and the torque $\tau$.  The color codes are the same as in \fig{fig:r027_T12msd}. Good isomorph invariance is generally observed except for significant short-time deviations for the two rotational autocorrelation functions (see the text).}
    \label{fig:r027_T12_VACF_AVACF}
\end{figure}

Next we show data for four time-autocorrelation functions. \Fig{fig:r027_T12_VACF_AVACF} gives in the two upper figures the velocity ($v$) and angular velocity ($\omega$) time-autocorrelation functions while the two lower figures give the force and torque time-autocorrelation functions. As in \fig{fig:r027_T12msd}, all functions are given in reduced units as functions of the reduced time $\tilde{t}$. Overall, we see in both the isotropic and the nematic phases good isomorph invariance and a sizable variation along the corresponding isochores. The short-time angular velocity and torque time-autocorrelation functions violate isomorph invariance significantly, however. This is due to the fact that the moment of inertia in the simulations was kept fixed, implying that this quantity is not constant in reduced units. As a consequence, the short-time ballistic motion is not isomorph invariant. At intermediate and long times, we do find good isomorph invariance also for the rotational time-autocorrelation functions; here the moment of inertia plays little role for the dynamics, which for a given molecule is dominated by interactions with the surrounding GB molecules. A weaker, but still clearly visible violation of isomorph invariance occurs at short times for the force autocorrelation function. In our understanding, this reflects the fact that the density-scaling exponent $\gamma$ changes with density along an isomorph, resulting in a changing effective inverse-power-law interaction. The lowest densities have the largest $\gamma$ (see the Appendix), leading to the highest average force squared coming from collisions. The collapse of the isochore angular velocity autocorrelation functions at short times is a consequence of the definition of reduced units, just as the short-time reduced-unit MSD collapse is.

\begin{figure}[!h]
    \includegraphics[width=0.45\textwidth]{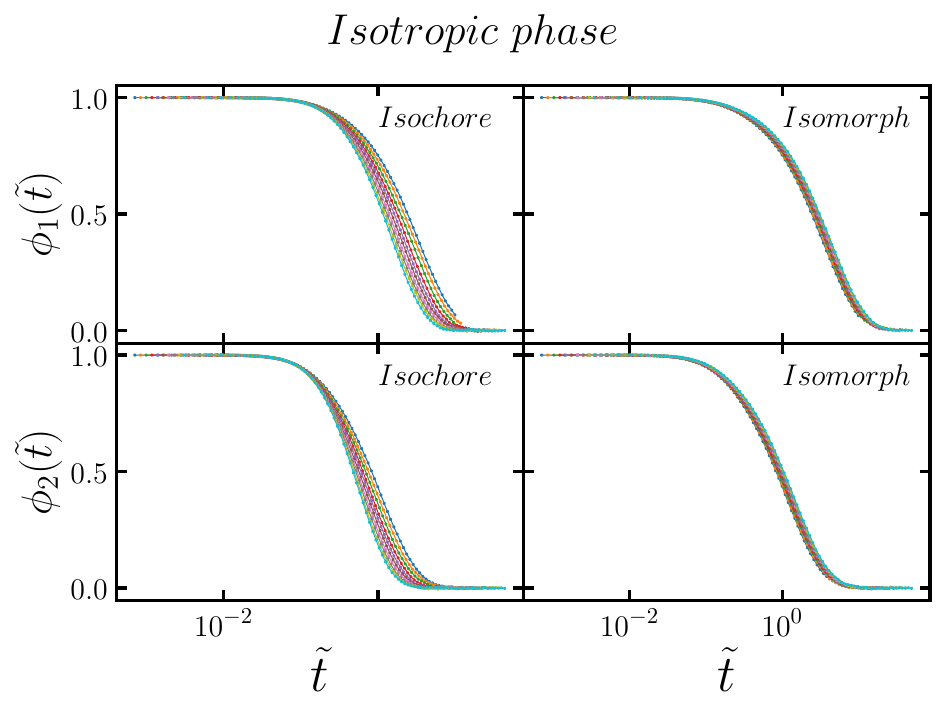}
    \includegraphics[width=0.45\textwidth]{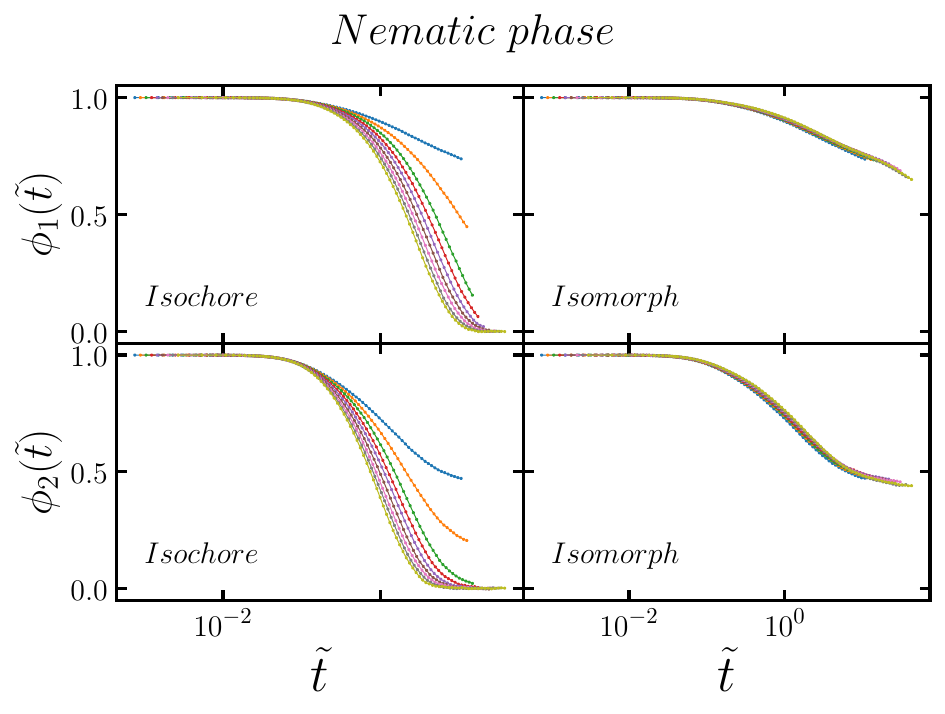}
    \caption{The left figures show the  first- and second-order orientational order parameter time-autocorrelation function in the isotropic phase, the right figures show the same in the nematic phase. Good isomorph invariance is observed in both phases.}
    \label{fig:r027_T12_OACF1_OACF2}
\end{figure}

Fig. \ref{fig:r027_T12_OACF1_OACF2} shows data for the first- and second-order orientational time-autocorrelation function, again plotted as functions of the reduced time. In the isotropic phase these functions go to zero at long times, confirming that there are no preferred orientations. This is not the case, of course, in the nematic phase. In both phases, we observe good isomorph invariance. According to the phase diagram (\fig{fig:heatmap}), the isochore defined from the nematic isomorph reference state point enters the isotropic phase at high temperatures; this is reflected in the figures by the fact that both time-autocorrelation functions go to zero at long times as the temperature increases. 

\begin{figure}[!h]
    \centering
    \includegraphics[width=0.45\textwidth]{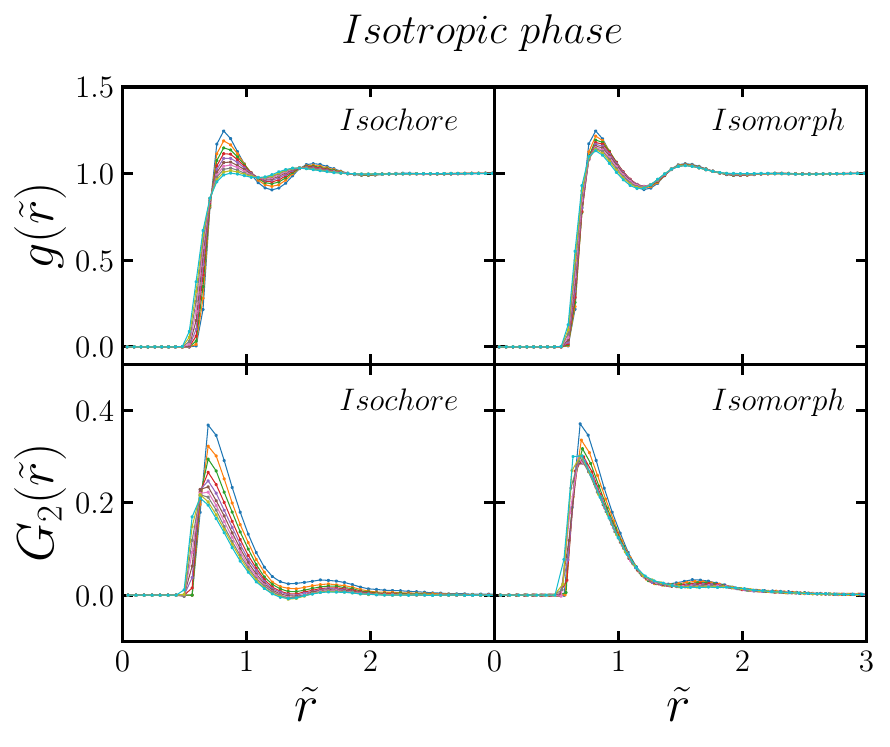}
    \includegraphics[width=0.45\textwidth]{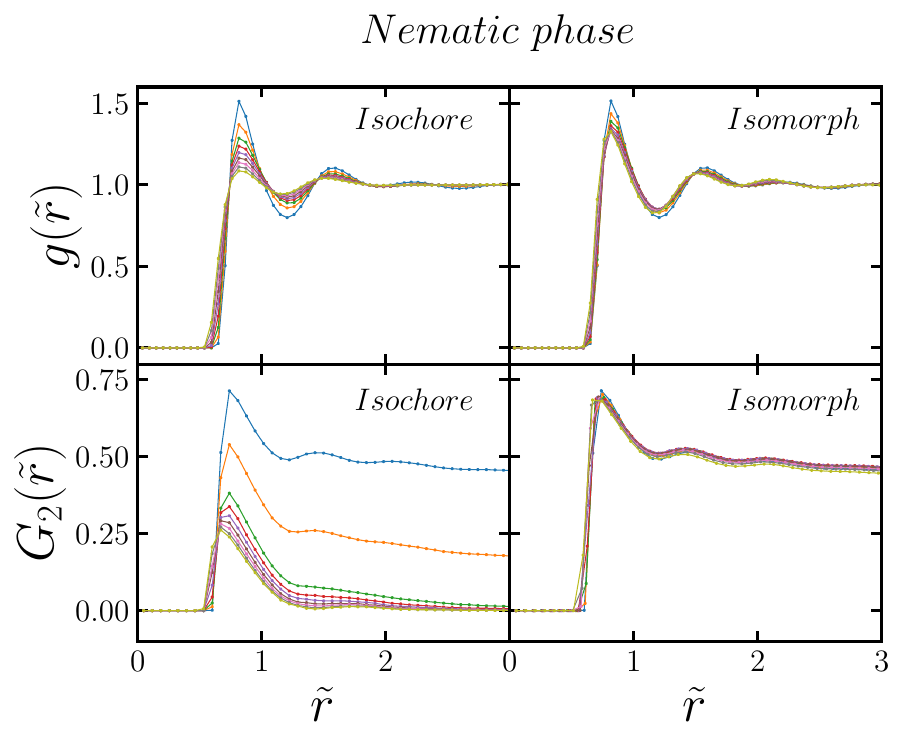}
    \caption{Reduced-unit radial distribution functions (upper figures) and radial-distribution orientational correlation function (\eq{G2}, lower figures) along the isochore and the isomorph in the isotropic (left) and nematic phases (right). Good isomorph invariance is observed in both phases, with some deviation at the first peak (see the text).}
    \label{fig:r027_T12_rdf_G2}
\end{figure}

So far we have discussed different dynamic signals and seen good isomorph invariance. The isomorph theory, however, also predicts that the reduced-unit structure should be invariant. This is tested in \fig{fig:r027_T12_rdf_G2}, in which the upper panels show the center-of-mass radial distribution function along the isochores and isomorphs. The data are close to invariant along the isomorph, though with visible deviations around the first peak. This is often observed when isomorph theory is tested over a large density range; it reflects a non-invariance arising when the density-scaling exponent $\gamma$ varies along the isomorph (similar to the non-invariance of the short-time force time-autocorrelation function discussed above). A large $\gamma$ implies a large effective inverse-power-law exponent, which decreases the probability of near contacts and is ``compensated'' by a higher peak in order to arrive at the same coordination number (defined by integration of the radial distribution function over its first peak). This phenomenon was recently rationalized in terms of isomorph invariance of the so-called bridge function of liquid-state theory \cite{cas21}. Confirming this interpretation, the highest first peaks along the isomorphs correspond to the lowest densities where $\gamma$ is largest (Appendix). 

Data for the orientational structure quantified in terms of $G_2(r)$ are given in the lower panels of \fig{fig:r027_T12_rdf_G2}. In the nematic phase, this function does not converge to zero at long times as in the isotropic phase. Nevertheless, except for the first peak in the isotropic phase, there is good isomorph invariance of the structure. Note that the tail of $G_2(r)$ goes to zero as the temperature is increased along the isochore in the nematic phase. This reflects a phase transition into the isotropic phase. In contrast, the tail is invariant as we increase the temperature along the isomorph.

\section{Conclusions}

When given in reduced units the dynamic and structural properties of the $GB(3,5,2,1)$ model are invariant to a good approximation along isomorphs in both the isotropic and the nematic phases, with some deviations at short times for orientational time-autocorrelation functions reflecting that the moment of inertia was assumed to be constant and thus not isomorph invariant in reduced units. In contrast, structure and dynamics are not invariant along isochores with the same temperature variation. Overall, our findings confirm isomorph-theory predictions and are consistent with the fact that the calamitic $GB(3,5,2,1)$ model obeys hidden scale invariance at high temperatures in both phases, i.e., has a virial potential-energy correlation coefficient above 0.9. For future work, it would be interesting to investigate the smectic phase of the model for which we based on \fig{fig:heatmap} expect good isomorphs even at temperatures lower than unity.

\begin{acknowledgments}
	This work was supported by the VILLUM Foundation's \textit{Matter} grant (16515).
\end{acknowledgments}
\newpage
\section*{Appendix: Isomorph state points}

This Appendix provides details of the two isomorphs studied, giving for each of these at several state points: density, temperature, virial potential-energy correlation coefficient $R$ (\eq{pearson}), and density-scaling exponent $\gamma$ (\eq{ds}).

\setlength{\tabcolsep}{30pt} 
\renewcommand{\arraystretch}{0.3} 
 \begin{table}[!htbp]
    \centering
    \resizebox{0.8\textwidth}{!}{%
	\centering
	\begin{tabular}{c  c  c  c}
	    \hline
		$\rho$ & $T$ & $R$ & $\gamma$ \\ [0.1ex] 
		\hline
		\hline
        0.2700 & 1.2000 &  0.9077 & 8.4553\\
        0.2727 & 1.3057 & 0.9158 & 8.5107\\
        0.2754 & 1.4215 & 0.9231 & 8.5535\\
        0.2782 & 1.5480 & 0.9285 & 8.5737\\
        0.2810 & 1.6859 & 0.9330 & 8.5795\\
        0.2838 & 1.8361 & 0.9367 & 8.5756\\
        0.2866 & 1.9995 & 0.9396 & 8.5615\\
        0.2895 & 2.1770 & 0.9416 & 8.5383\\
        0.2924 & 2.3697 & 0.9433 & 8.5126\\
        0.2953 & 2.5788 & 0.9445 & 8.4820\\
        0.2982 & 2.8056 & 0.9457 & 8.4555\\
        0.3012 & 3.0514 & 0.9461 & 8.4252\\
        0.3042 & 3.3177 & 0.9467 & 8.3962\\
        0.3073 & 3.6062 & 0.9469 & 8.3646\\
        0.3104 & 3.9186 & 0.9468 & 8.3364\\
        0.3135 & 4.2570 & 0.9469 & 8.3084\\
        0.3166 & 4.6231 & 0.9465 & 8.2815\\
        0.3198 & 5.0194 & 0.9463 & 8.2542\\
        0.3230 & 5.4483 & 0.9460 & 8.2276\\
        0.3262 & 5.9125 & 0.9456 & 8.2032\\
        0.3295 & 6.4148 & 0.9450 & 8.1802\\
        0.3327 & 6.9583 & 0.9446 & 8.1647\\
        0.3361 & 7.5463 & 0.9442 & 8.1479\\
        0.3394 & 8.1827 & 0.9436 & 8.1289\\
        0.3428 & 8.8712 & 0.9428 & 8.1121\\
        0.3463 & 9.6164 & 0.9423 & 8.0987\\
        0.3497 & 10.4226 & 0.9416 & 8.0867\\
        0.3532 & 11.2953 & 0.9410 & 8.0735\\
        0.3567 & 12.2396 & 0.9403 & 8.0628\\
        0.3603 & 13.2616 & 0.9396 & 8.0562\\
        0.3639 & 14.3680 & 0.9390 & 8.0462\\
        0.3676 & 15.5657 & 0.9383 & 8.0443\\
        0.3712 & 16.8622 & 0.9376 & 8.0396\\
        0.3749 & 18.2660 & 0.9370 & 8.0345\\
        0.3787 & 19.7859 & 0.9362 & 8.0315\\
        0.3825 & 21.4318 & 0.9356 & 8.0285\\
        0.3863 & 23.2145 & 0.9350 & 8.0295\\
        0.3902 & 25.1457 & 0.9342 & 8.0295\\
        0.3941 & 27.2379 & 0.9335 & 8.0343\\\\
		\hline
	\end{tabular}
	}
	\caption{Variation of density, temperature, correlation coefficient $R$, and density-scaling exponent $\gamma$ for the state points on the isotropic-phase isomorph generated from the reference state point $(\rho_{\rm ref},T_{\rm ref})=(0.27,1.2)$.}
	\label{table:r0.27_T1.2}
 \end{table}
 
 \setlength{\tabcolsep}{30pt} 
\renewcommand{\arraystretch}{0.3} 
 \begin{table}[!htbp]
    \centering
    \resizebox{0.8\textwidth}{!}{%
	\centering
	\begin{tabular}{c  c  c  c}
	    \hline
		$\rho$ & $T$ & $R$ & $\gamma$ \\ [0.1ex] 
		\hline
		\hline
        0.3300 & 1.2000 & 0.9169 & 8.2217\\
        0.3333 & 1.3027 & 0.9230 & 8.2809\\
        0.3366 & 1.4149 & 0.9274 & 8.3187\\
        0.3400 & 1.5371 & 0.9315 & 8.3349\\
        0.3434 & 1.6700 & 0.9342 & 8.3301\\
        0.3468 & 1.8142 & 0.9359 & 8.3183\\
        0.3503 & 1.9705 & 0.9375 & 8.2957\\
        0.3538 & 2.1398 & 0.9385 & 8.2664\\
        0.3573 & 2.3229 & 0.9389 & 8.2333\\
        0.3609 & 2.5209 & 0.9393 & 8.2066\\
        0.3645 & 2.7349 & 0.9394 & 8.1697\\
        0.3682 & 2.9660 & 0.9392 & 8.1395\\
        0.3719 & 3.2156 & 0.9393 & 8.1057\\
        0.3756 & 3.4851 & 0.9387 & 8.0731\\
        0.3793 & 3.7759 & 0.9384 & 8.0433\\
        0.3831 & 4.0898 & 0.9375 & 8.0058\\
        0.3870 & 4.4283 & 0.9367 & 7.9788\\
        0.3908 & 4.7934 & 0.9362 & 7.9487\\
        0.3947 & 5.1873 & 0.9354 & 7.9212\\
        0.3987 & 5.6119 & 0.9345 & 7.8922\\
        0.4027 & 6.0697 & 0.9337 & 7.8677\\
        0.4067 & 6.5632 & 0.9321 & 7.8438\\
        0.4108 & 7.0953 & 0.9315 & 7.8229\\
        0.4149 & 7.6689 & 0.9306 & 7.7996\\
        0.4190 & 8.2870 & 0.9296 & 7.7791\\
        0.4232 & 8.9533 & 0.9286 & 7.7620\\
        0.4274 & 9.6715 & 0.9277 & 7.7464\\
        0.4317 & 10.4455 & 0.9264 & 7.7286\\
        0.4360 & 11.2797 & 0.9258 & 7.7155\\
        0.4404 & 12.1786 & 0.9244 & 7.6942\\
        0.4448 & 13.1473 & 0.9232 & 7.6844\\
        0.4492 & 14.1912 & 0.9223 & 7.6712\\
        0.4537 & 15.3160 & 0.9209 & 7.6535\\
        0.4583 & 16.5277 & 0.9197 & 7.6442\\ \\
		\hline
	\end{tabular}
	}
	\caption{Variation of density, temperature, correlation coefficient $R$, and density-scaling exponent $\gamma$ for the state points on the nematic-phase isomorph generated from the reference state point $(\rho_{\rm ref},T_{\rm ref})=(0.33,1.2)$. }
	\label{table:r0.33_T1.2}
 \end{table}

\end{document}